# Duplex Mikaelian lenses and duplex Maxwell's fish eye lenses

Huiyan Peng [1, 2, 3], Senlin Liu [1, 2], Yuze Wu [1, 2], Yi Yan [1, 2], Zichun Zhou [1, 2], Xiaochao Li [1, 2], Qiaoliang Bao [4], Lin Xu [3, †], and Huanyang Chen[1, 2, ‡]

[1] Institute of Electromagnetics and Acoustics and Key Laboratory of Electromagnetic Wave Science and Detection Technology, Xiamen University Xiamen 361005, China

[2] Department of Electrical and Electronics Engineering, Xiamen University Malaysia, 43900 Sepang, Selangor, Malaysia

[3] Institutes of Physical Science and Information Technology and Key Laboratory of Opto-Electronic Information Acquisition and Manipulation of Ministry of Education, Anhui University, Hefei 230601, China

[4] Department of Materials Science and Engineering, and ARC Centre of Excellence in Future Low-Energy Electronics Technologies (FLEET), Monash University, Clayton, Victoria 3800, Australia

† E-mail: xuin@ahu.edu.cn
‡ E-mail: kenyon@xmu.edu.cn

In this paper, we report two new kinds of absolute optical instruments that can make stigmatically images for geometric optics in two dimensional space. One is called the duplex Mikaelian lens, which is made by splicing two half Mikaelian lenses with different periods. The other is exponential conformal transformer of duplex Mikaelian lens with the ratio of different periods of its two half Mikaelian lenses a rational number, which we call duplex Maxwell's fish eye lens. Duplex Mikaelian lenses have continuous translation symmetry with arbitrary real number, while duplex Maxwell's fish eye lenses have continuous rotational symmetry from $0$ to $2\pi$. Hence each duplex Maxwell's fish eye lens corresponds to a duplex Mikaelian lens. We further demonstrate the caustic effect of geometric optics in duplex Mikaelian lenses and duplex Maxwell's fish eye lenses. In addition, we investigate the Talbot effect of wave optics in the duplex Mikaelian lens based on numeric calculations. Our findings based on splicing and exponential conformal mapping enlarge the family of absolute optical instruments.

## I . Introduction

Recently, absolute optical instruments (AOIs) [1-7] that can make stigmatically images in geometrical optics have drawn a lot of attention, such as Maxwell's fish eye lens[1], Luneburg lens[3], and Miñano lens[4]. They are widely used in different areas, like subwavelength focusing [8-10] and transformation optical designs [11-13]. Two-dimensional (2D) AOIs with gradient refractive index of rotational symmetry can be casted into an another 2D AOIs with gradient refractive index of continuous translation symmetry by exponential conformal mapping $w = \exp(z)$. The well-studied example is the pair of a Maxwell's fish eye lens and a Mikaelian lens [14]. The gradient refractive index of Mikaelian lens is written as $1/\cosh(x)$ [15]. Besides the continuous translation symmetry along $y$ axis, Mikaelian lens has discrete mirror symmetry of $y$ axis. Self-Focusing of geometrical optics and Talbot effect of wave phenomena in Mikealian lens had been investigated both theoretically and experimentally [14].

In this paper, we construct two new kinds of AOIs. One is made by splicing two half Mikaelian lenses with different periods, which we shall call the duplex Mikaelian lens.

The duplex Mikaelian lens has continuous translation symmetry in one direction without mirror symmetry in another direction. By exponential conformal mapping, the duplex Mikaelian lens with the ratio of different periods of its two half Mikaelian lenses a rational number can be casted into another kind of AOI with continuous rotational symmetry, which we call duplex Maxwell's fish eye lens. This paper is organized as follows. In Sec II, we introduce Mikaelian lenses and their exponential conformal transformers. In Sec. III, we construct the duplex Mikaelian lens and the duplex Maxwell's fish eye lens. In Sec. IV, we demonstrate the caustic effect of geometric optics in duplex Mikaelian lenses and duplex Maxwell's fish eye lenses. In Sec. V, the Talbot effect of wave optics in the duplex Mikaelian lens is illustrated based on numeric calculations. In Sec.VI, we give a conclusion.

**II . Mikaelian lenses and their exponential conformal transformers**

In general, Mikaelian lens in two dimensional space can be expressed in,

$$n(x) = 1/\cosh(p \cdot x/a), \qquad (1)$$

where $a$ is a reference scale length and $p$ is a positive real number. In principle, we can absorb $p$ into $a$ and redefine $a/p$ as $a'$, therefore there could be no $p$ term in Eq. (1). But for following discussion of exponential conformal mapping, we call the lens with gradient refractive index satisfying Eq. (1) the Mikaelian lens with $\{p\}$. The continuous translation symmetry of Eq. (1) is of one dimensional Lie group R$^1$, namely, real numbers, which is non-compact. The light rays marked with red and blue lines from a red point source with the position coordinate $(0,0)$ can make self-focusing, see the red and blue points with the interval $\pi a/2$ along $y$ axis in Fig. 1(a) with $p=2$. As it is illustrated, during the reference period length $L = 2\pi a$ between two horizontal dashed lines, we have two periods for Mikaelian lens with refractive index written as Eq. (1). Hence it is a Mikaelian lens with the period length of $L_0 = L/p$, where light rays travel periodically. The light trajectories in the Mikaelian lens are expressed by $k*Sinh(x/a) = Sin(y/a)$ [15], where $k$ is the slope of the light ray at the red point source with the position coordinate $(0,0)$. In the following paper, we perform the ray tracing calculation based on Hamtionian equation of optics [16].

Now, let us revisit the exponential conformal mapping [14] written as,

$$w/a = \exp(z/a), \quad z/a = \log(|w/a|) + (\arg(w/a) + 2j\pi)i, \qquad (2)$$

where $j$ is an integer. It connects a Riemann surface as shown by the chiral rotated surface (denoted with $w = u + vi$) in Fig. 1(b) to a complex plane (denoted with $z = x + yi$) in Fig. 1(a). The chiral rotated surface is the schematic plot of the Riemann surface, which looks like the wave front of right circular polarized light. Each region (marked with $j$) between dashed lines is indeed a plane with a branch cut (see the dashed line) as shown in Fig. 1(c). Here we use the chiral rotated surface to lift the third space dimension for a better illustration of branch cuts, which will be useful for further discussion. One important thing is that each ribbon region between dashed lines marked with the integer $j$ in Fig. 1(a) has the reference period length

$L = 2\pi a$, which corresponds to a complex plane. The $y$ axis denoted by the black line in Fig. 1(a) is mapped to a spiral black curve in Fig. 1(b), hence corresponding to the black circle in Fig. 1(c). According to conformal transformation optics [11, 17], the gradient refractive index between $z$-space and $w$-space can be expressed by

$$n(w) = |dz/dw| n(z), \qquad (3)$$

where $n(z)$ and $n(w)$ are the gradient refractive indexes of $z$-space and $w$-space respectively. Hence, the gradient refractive index $n(z) = n(x+yi) = n(x)$ in Eq. (1) of Mikaelian lens with $\{p\}$ in $z$-space can be casted into a gradient refractive index in $w$-space with the form of

$$n(w) = 2/\left(abs(w/a)^{1+p} + abs(w/a)^{1-p}\right), \qquad (4)$$

whose continuous rotational symmetry is of another one dimensional Lie group $S^1$, namely circle group from $0$ to $2\pi$, which is compact comparing to Lie group $R^1$. When $p$ is a rational number, Eq. (4) denotes the gradient refractive index of the generalized Maxwell's fish eye lens [18], which we will call it the Maxwell's fish eye lens with $\{p\}$. The conventional Maxwell's fish eye lens refers to Maxwell's fish eye lens with $\{p=1\}$, which is an AOI in optics as a purely geometrical device [5]. The light rays in such a Maxwell's fish eye lens with $\{p\}$ are all closed, as shown in an example with blue and red curves in Fig. 1(c). The red point source at the position coordinate $(1,0)$ can make other three images in the Maxwell's fish eye lens with $\{p=2\}$. The corresponding trajectories are shown in blue and red spiral curves of light rays in Riemann surface in Fig. 1(b). As it is also shown in the following Fig. 3(d), a point source in the Maxwell's fish-eye lens with $\{p=1\}$ only has one image.

The Mikaelian lens with any positive real number $\{p\}$ in Eq. (1) represents an AOI, whose period length is $L_0 = L/p$. After exponential conformal mapping we obtain a new lens with gradient refractive index in Eq. (4). Why only $p$ is rational number in Maxwell's fish eye lens that can make it an AOI? We will give a heuristic discussion based on its corresponding Mikaelian lens by exponential conformal mapping in Eq. (2). Suppose that the rational number of Maxwell's fish eye lens with $\{p\}$ can be denoted by $p = p_u / p_d$, where $p_u$ and $p_d$ are coprime. Now its corresponding Mikaelian lens has the period length of $L_0 = L p_d / p_u$. In this case, we can always find a large period $L_g = L p_d = L_0 p_u$, which is $p_d$ times of the reference period length $L$ and $p_u$ times of the period length $L_0$ of the Mikaelian lens. Therefore, in the region of large period $L_g$, light rays from one point source have $p_d$ periods (or $2p_d$ focusing points) in the Mikaelian lens, which corresponds to wind around the original point of Maxwell's fish eye lens $p_u$ times and then recover to the original propagation condition. To be more concrete, we plot Mikaelian lens with $\{p_u / p_d = 2/3\}$ in Fig. 1(d) and its corresponding Maxwell's fish eye lens in Fig. 1(f).

As we can see, light rays from one red point source have $2$ periods (or $4$ focusing points) in the Mikaelian lens of Fig. 1(d) during the large period $L_g = 3L$ between two purple solid lines. In the corresponding Riemann surface of Fig. 1(e), light rays can wind around the center axis 3 times and then make perfect imaging. When we consider light rays in real Maxwell's fisheye lens in Fig. 1(f) projected from Riemann surface, there still exists perfect imaging. Once $p$ is not a rational number [19], there is no way to find a large period, or to say the large period is infinite. Hence light rays will wind the center axis infinite times without recovering their original propagation condition, which means that Maxwell's fish eye lens with irrational number index is not an AOI. However, we know that Mikaelian lens with any positive real number index is an AOI that can make self-focusing. The difference comes from Lie group $R^1$ of Mikaelian lens and Lie group $S^1$ of Maxwell's fisheye lens.

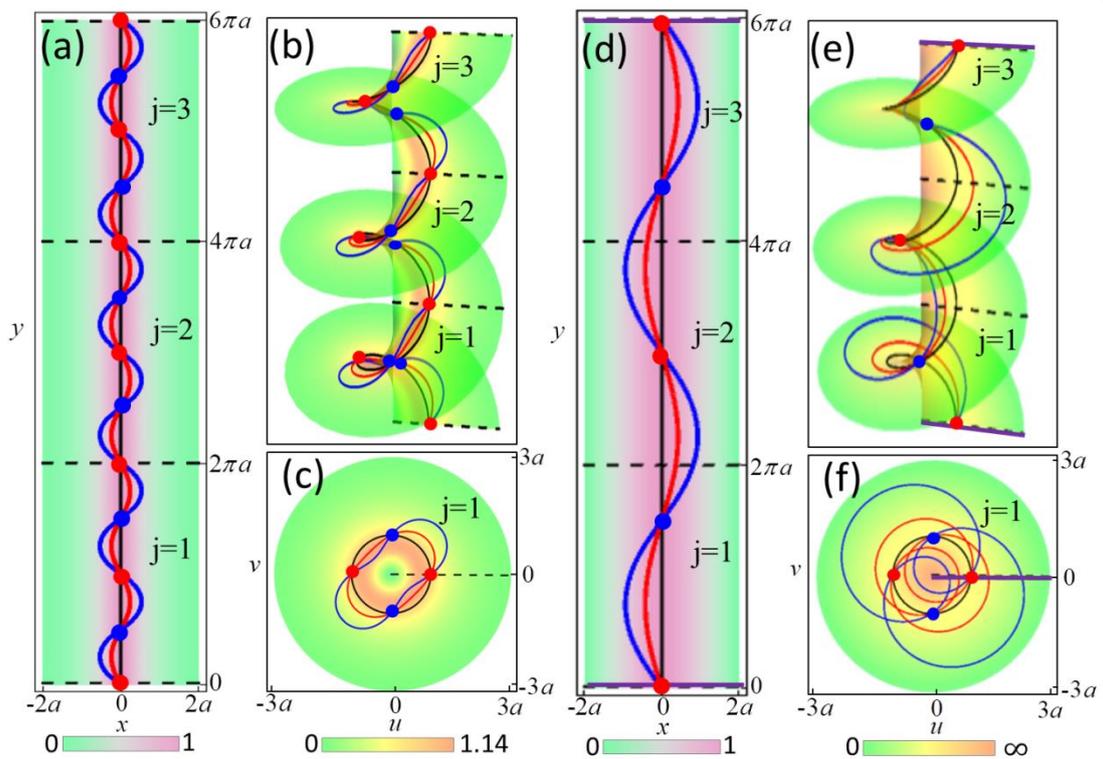

Fig. 1  The gradient refractive index and light trajectories in the Mikaelian lens $\{p\}$ and the corresponding Maxwell's fish eye lens. (a) The Mikaelian lens with $\{p=2\}$ is in $z$-space, which is divided into infinite number ribbon regions (marked with integer $j$) with a reference period length $L = 2\pi a$ between dashed lines. Its gradient refractive index written as Eq. (1) is shown with the contour plot in complex plane $z = x + yi$. Two red and blue light rays launch from the red source at the position coordinate $(0,0)$. In the reference period length, the red source can make three more images marked with two blue points and a red point; (b) The schematic plot of Riemann surface of exponential conformal mapping. Branch cuts are denoted by dashed lines, which are mapped from the dashed lines of $z$-space. The chiral rotated surface is used to lift a third space dimension for a better illustration of branch cuts; (c) The corresponding gradient refractive index and light trajectories in the Maxwell's fish eye lens, which is mapped from the ribbon region marked with $j=1$ of the left figure by exponential conformal mapping. Indeed, it is also the projection from multiple $w$-space of the Riemann surface of (b). The radius $a$ denoted with black circle is mapped from $y$ axis of $z$-space; (d) The Mikaelian lens with $\{p=2/3\}$ is in $z$-space. Two additional purple lines denote boundaries of the large period; (e) Light rays in the Riemann surface of Maxwell's fish eye lens with $\{p=2/3\}$. (d) Light rays in Maxwell's fish eye lens with $\{p=2/3\}$.

**III. Duplex Mikaelian lens and duplex Maxwell's fish eye lens**

Since the gradient refractive index of different Mikealian lenses along $y$ axis are all 1, we splice two half Mikaelian lenses with different real number index $\{p\}$ and $\{q\}$ to make a duplex Mikaelian lens with $\{p,q\}$, see the half Mikaelian lens with $\{p=2\}$ on the left and the half Mikaelian lens with $\{q=1\}$ on the right in Fig. 2(a). It is clear that the gradient refractive index of a duplex Mikaelian lens with $\{p,q\}$ does not have discrete mirror symmetry of $y$ axis. The derivative of gradient refractive index of Mikaelian lens is written as,

$$\partial_x n(x) = -\text{sech}(px/a)\tanh(px/a)p/a, \quad (5)$$

where its value at $y$ axis is zero. Therefore, the gradient refractive index and its derivative of duplex Mikaelian lens are both continuous. When light rays pass thought $y$ axis, they can be welded together smoothly as shown by the focusing points of $y$ axis in Fig. 2(a). Each half of Mikaelian lens contributes half period to the duplex Mikalian lens. Hence the period of the duplex Mikaelian lens with $\{p,q\}$ is $L_D = L/2/p + L/2/q$ along the $y$ axis, which is no long the period of Mikaelian lens with $\{p\}$ or that with $\{q\}$. Such duplex Mikaelian lens with $\{p,q\}$ can also make self-focusing effect as shown in Fig. 2(a), which is a new kind of AOIs. There are infinitely many combinations of duplex Mikaelian lenses with asymmetric gradient refractive indexes.

Now comes the question, is exponential conformal transformer of duplex Mikaelian lens with any two different real number $\{p,q\}$ an AOI? The answer is no. Only exponential conformal transformer of duplex Mikaelian lens with two rational number $\{p,q\}=\{p_u/p_d, q_u/q_d\}$ ( $p_u, p_d$, $q_u$ and $q_d$ are all coprime ) is an AOI. We would like call such transformer the duplex Maxwell's fish eye lens with rational number $\{p,q\}$. Let's explain the reason based on exponential conformal mapping in Eq. (2). From the previous discussion, it would be much easier if we start from duplex Mikaelian lens with $\{p_u/p_d, q_u/q_d\}$. Its period $L_D = \pi a p_d/p_u + \pi a q_d/q_u$ and the reference period length $L=2\pi a$ are two important lengths for the property of duplex Mikaelian lenses. One can find a large period $L_{Dg} = (p_u q_d + p_d q_u)L = 2p_u q_u L_D$, which is $p_u q_d + p_d q_u$ times of $L$ and $2p_u q_u$ times of $L_D$. Therefore, during the region of a large period $L_{Dg}$, light rays from one point source have $2p_u q_u$ periods in the duplex Mikaelian lens, which corresponds to wind around the original point of its exponential conformal transformer $p_u q_d + p_d q_u$ times and then recover to the original propagation condition.

Let us see an example in Fig. 2(a) and 2(b), $\{p = p_u/p_d = 2/1\}$ and $\{q = q_u/q_d = 1/1\}$. In the region of a large period $L_{Dg}$ between two purple solid lines, there are 3 periods of the reference lengths and 4 periods of duplex Mikaelian lens. Such 3 periods of the reference lengths can be mapped to the Riemann surface between two purple solid lines in Fig. 2(b), which can be further projected to one complex plane in Fig. 2(c). Light rays in Fig. 2(b) and Fig. 2(c) winds 3 times around the center, while they travel back and forth along the black line 4 times. In this way, such duplex Maxwell's fish eye lens with rational number indexes can make perfect imaging, which makes them AOIs. The light rays from a red point source at the position coordinate $(0,0)$ can make self-focusing in such a duplex Mikaelian lens along $y$ axis, see blue and red lines in Fig. 2(a). It should be noticed that light rays from a point source located at any point can make one more image during one period of a Mikaelian lens. However, in a duplex Mikaelian lens, light rays from a point source off $y$ axis cannot make other image during its period (we will show it in the next section about the caustic effect). After the exponential conformal mapping, the duplex Mikaelian lens becomes a new type of AOI as shown in Fig. 2(c), which is the projection of Riemann surface of Fig. 2(b) to a complex plane. This new type of AOI is a duplex Maxwell's fish eye lens with two rational numbers $\{p,q\}$, where $p$ is the index inside the circular region of radius $a$ and $q$ outside it. Perfect imaging will also be valid in this lens, as shown by the red and blue closed trajectories and focusing points in Fig. 2(c). If two half Mikaelian lenses in Fig. 2(a) are exchanged with their positions, we can obtain another duplex Mikaelian lens with two rational numbers $\{q, p\}$ in Fig. 2(d). Its exponential conformal transformer is shown in Fig. 2(e) and 2(f). Although the duplex Mikaelian lens with $\{p, q\}$ and duplex Mikaelian lens with $\{q, p\}$ are similar under a mirror reflection (see examples between Fig. 2(a) and 2(d)), but their exponential conformal transformers are quite different in their gradient refractive indexes (see examples between Fig. 2(c) and 2(f)). As for duplex Mikaelian lens without two rational number index, there is no way to find a large period, hence there is no corresponding duplex Maxwell's fish eye lens. Therefore, we have two new kinds of AOIs, which haven't been proposed as far as we know. One is the duplex Mikaelian lens with any two different positive real number $\{p, q\}$, the other is the duplex Maxwell's fish eye lens with rational number $\{p, q\}$.

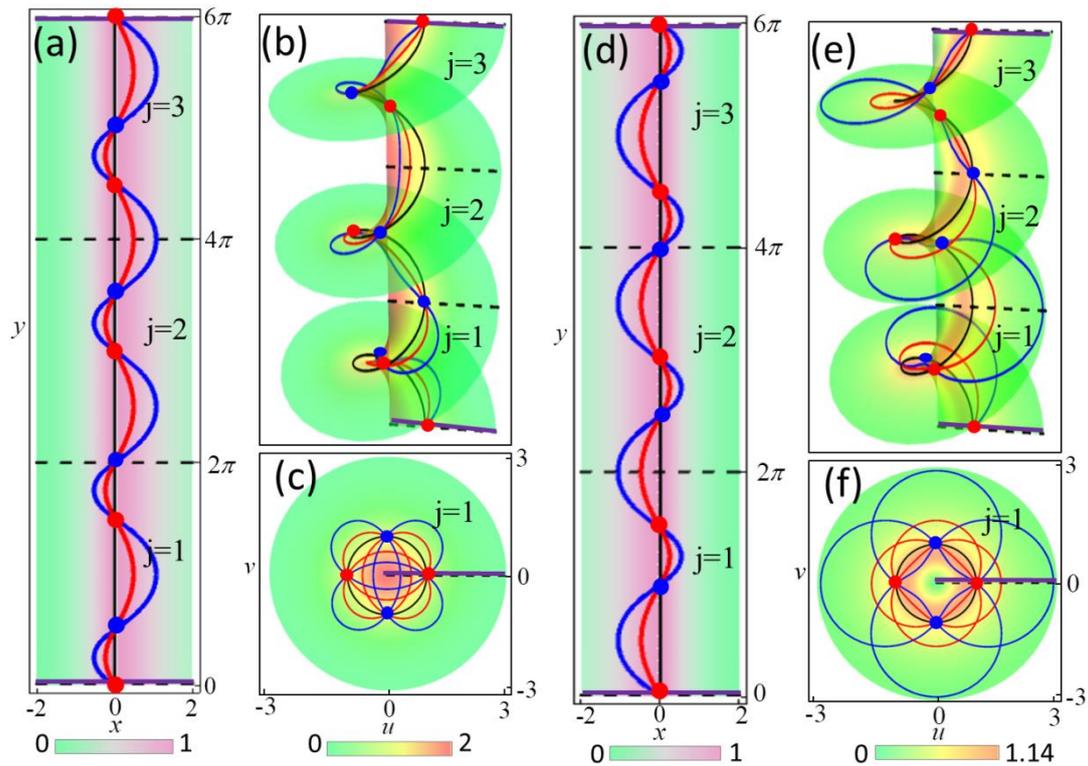

Fig. 2. The gradient refractive index and light trajectories in the duplex Mikaelian lens with $\{p,q\}$ and the corresponding duplex Maxwell's fish eye lens. (a) The Mikaelian lens with $\{2,1\}$ is in $z$-space. Its gradient refractive index is without mirror symmetry along $y$ axis as shown with contour plot. During the reference period length $L = 2\pi a$, two red and blue light rays launch from the red source at the position coordinate $(0,0)$ region marked with $j=1$ of the left figure, which can make two more images; (b) The schematic plot of Riemann surface and light trajectories of the duplex Maxwell's fish eye lens with $\{2,1\}$; (c) The projection to $w$-space of the Riemann surface of (b); (d) The Mikaelian lens with $\{1,2\}$ is in $z$-space; (e) The schematic plot of Riemann surface and light trajectories of the duplex Maxwell's fish eye lens with $\{1,2\}$; (f) The projection to $w$-space of the Riemann surface of (e).

**IV. Caustics in duplex Mikaelian lenses and duplex Maxwell's fish eye lenses**

In this section, we will systematically investigate the property of duplex Mikaelian lens and duplex Maxwell's fish eye lens in geometric optics. For a Mikaelian lens with $\{p=1\}$, there will be two focusing points for a reference period of $L$, as shown in Fig. 3(a). For a better illustration, we rotate clockwise the figure by $\pi/2$ corresponding to Fig. 1(a). While for a Mikaelian lens with $\{p=2\}$, there will be four focusing points for a reference period of $L$, as shown in Fig. 3(b). After merging into a duplex Mikaelian lens with $\{p=1, q=2\}$, there will be one focusing point and one caustic curve in a period of duplex Mikaelian lens shown in Fig. 3(c), which is also mentioned in Sec. III. The caustic of optics can be defined as the envelope of a system of orthotomic rays [20-23]. In Fig. 3(c), there exist caustic curves marked with blue dashed lines, where light will be converged along a particular curve. If the red source is approaching to the $y$ axis, the blue dashed caustics are becoming smaller as well approaching to the $y$ axis. Once the red source is on the $y$ axis just as shown in Fig. 2(a), the blue dashed caustics are shrunk to blue points. Those caustics can be treated as reminiscences of imaging points in the other side of $y$ axis due to the asymmetric gradient refractive index. Such a caustic effect will be very obvious in the corresponding duplex Maxwell's fish eye lens after exponential conformal mapping. For the Maxwell's fish eye lens with $\{p=1\}$, the colored light rays from a source point (in red) at the position coordinate $(1.5a, 0)$ outside radius $a$ can make a perfect image point (in blue) inside, as shown in Fig. 3(d). Include the source (or self-image) point, there will be four imaging points (in red and blue) in the

Maxwell's fish eye lens with $\{p=2\}$ (see in Fig. 3(e)). After the combination, there will be four images (in red) outside radius $a$, as shown in Fig. 3(f). The image number is exactly the number of $2p_u q_u$ (here 4). The four caustic curves in Fig. 3(c) are now mapped to square-like caustic curve inside radius $a$ of Fig. 3(f). If the red source is approaching to radius $a$ as shown in Fig. 2(f), the blue dashed caustics are shrunk to blue points, which will overlap the red points.

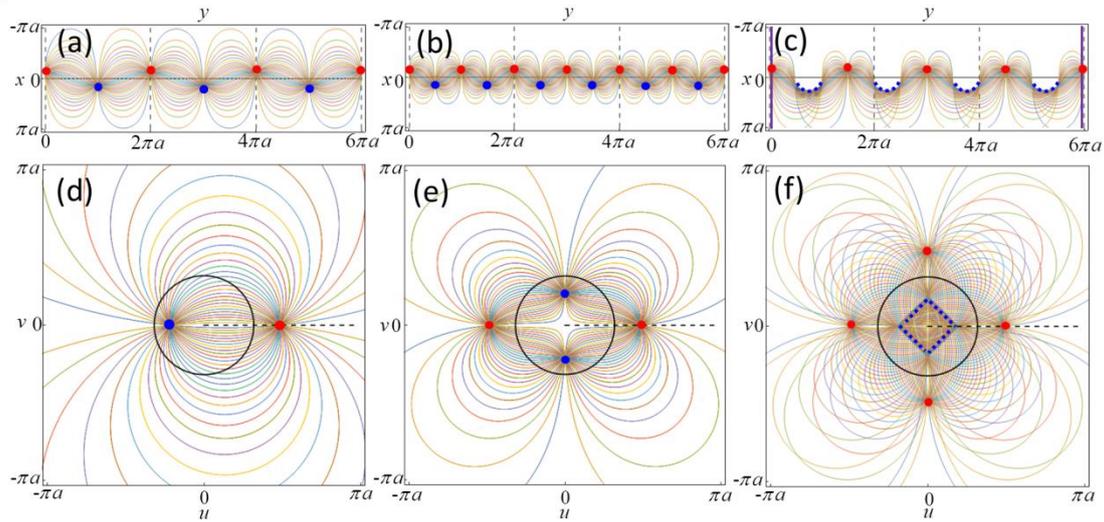

Fig. 3 (a) Colored light rays in the Mikaelian lens with $\{p=1\}$ are emitted from and focused on red and blue points, which are off $y$ axis; (b) Colored light rays in the Mikaelian lens with $\{p=2\}$ are emitted from and focused on red and blue points, which are off $y$ axis; (c) Colored light rays in the duplex Mikaelian lens with $\{p=2, q=1\}$ are emitted from and focused on red points, which are off $y$ axis. The blue dashed curves denote caustic curves; (d) Colored light rays in the Maxwell's fish eye lens with $\{p=1\}$ are emitted from and focused on red and blue points, which are off radius $a$; (e) Colored light rays in the Maxwell's fish eye lens with $\{p=2\}$ are emitted from and focused on red and blue points, which are off radius 1; (f) Colored light rays in the duplex Maxwell's fish eye lens with $\{p=2, q=1\}$ are emitted from and focused on red points, which are off radius $a$. The blue dashed square-like curves denote caustic curves.

To enhance the visualization the above effect, we consider a duplex Maxwell's fish eye lens with $\{p=1, q=3\}$ in Figs. 4(a), 4(b), and 4(c). In Fig. 4(a), the light rays from the point source (in red) with the position coordinate $(1.5a, 0)$ will generate two more red images. Inside the radius $a$, there exists a triangular-like caustic curve. While the red point source is approaching to radius $a$, the triangular-like caustic curve shrinks to three blue points, as shown in Fig. 4(b). Therefore, it has six images

(three red points and three blue points) at radius $a$. When the red light source is placed with the position coordinate $(0.5a,0)$ inside radius $a$, the triangular caustic curve locates outside, see the blue dashed curve in Fig. 4(c). The evolution of caustic curve in a duplex Maxwell's fish-eye lens with $\{p=3, q=1\}$ is plotted in Figs. 4(d), 4(e) and 4(f). In Fig. 4(d), the red point source is located at with the position coordinate $(2a,0)$, and there will be two more red images outside radius $a$, a clover-shaped caustic curve inside radius $a$. After moving the red point source to radius $a$ as shown in Fig. 4(e), the clover-shaped caustic curve shrinks to blue points. While the red source point placed at $(0.8a,0)$, the clover-shaped caustic curve locates outside, as shown in Fig. 4(e).

In fact, as long as the ratio of $p$ and $q$ is a rational number, the imaging and caustic effect are still valid. However, if the ratio becomes irrational number, there will be no perfecting imaging, similar to the effect found in Ref. [19]. Hence the caustic effect disappears.

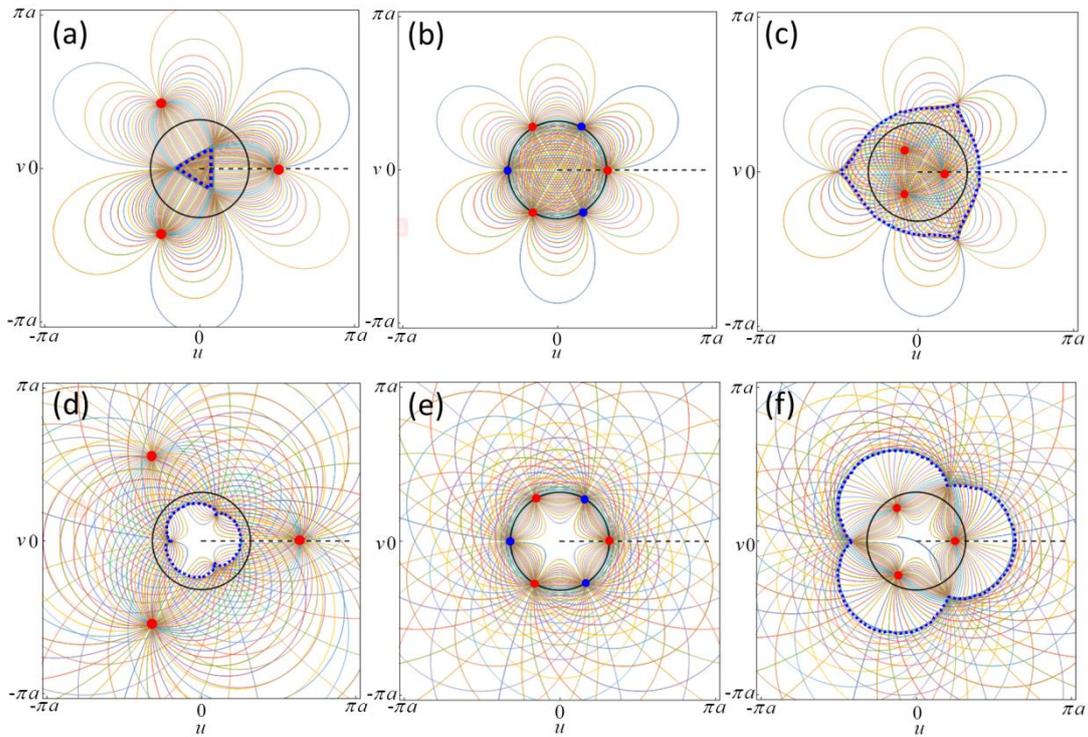

Fig. 4 (a) Colored light rays in the duplex Maxwell's fish eye lens with $\{p=3, q=1\}$ are emitted from and focused on red points, which are outside radius $a$. The blue dashed triangular-like curve inside radius $a$ is caustic curve; (b) Colored light rays in the duplex Maxwell's fish eye lens with $\{p=3, q=1\}$ are emitted from and focused on red and blue points, which are on radius $a$; (c) Colored light rays in the duplex Maxwell's fish eye lens with $\{p=3, q=1\}$ are emitted from and focused on red points, which are inside radius $a$. The blue dashed triangular-like curve outside radius $a$ is caustic curve; (d) Colored light rays in the duplex Maxwell's fish eye lens with $\{p=1, q=3\}$ are emitted from and focused on red points, which are outside radius $a$. The blue dashed clover-shaped curve inside radius $a$ is caustic curve; (e) Colored light rays in the duplex Maxwell's fish-eye lens with $\{p=1, q=3\}$ are emitted from and focused on red and blue points, which are on radius $a$; (f) Colored light rays in the duplex Maxwell's fish eye lens with $\{p=1, q=3\}$ are emitted from and focused on red points, which are inside radius $a$. The blue dashed clover-shaped curve outside radius $a$ is caustic curve;

**V. Talbot effect in duplex Mikaelian lenses**

In this section, we will focus on the property of duplex Mikaelian lens in wave optics. The following full-wave simulations are performed by the RF module of commercial software COMSOL Multiphysics. It has been demonstrated theoretically and experimentally that Talbot effect is existed in the Mikaelian lens[14]. Talbot effect is the self-imaging of a grating in optical system when it is illuminated by a monochromatic plane wave [24-27]. In Fig 5(a), array sources with their electric component (with unit V/m) out of plane of electromagnetic wave (with frequency 1THz, $a=1mm$) satisfying $abs(1-y)$ are placed in metallic grating along $x$ axis of the Mikaelian lens with $\{p=2\}$. Here we also rotate clockwise the figure by $\pi/2$. We plot their electric component at every half period $L_0/2=\pi$ in Fig. 5(c). It shows the Talbot effect of wave phenomenon, where the image pattern of array sources can reappear at every half period. For the duplex Mikaelian lens with $\{p=2, q=1\}$ of Fig. 2(a), we plot the electric field in Fig. 5(b), which is under the same metallic grating source setting of Fig. 5(a). The electric component at every period $L_D$ is plot in Fig.

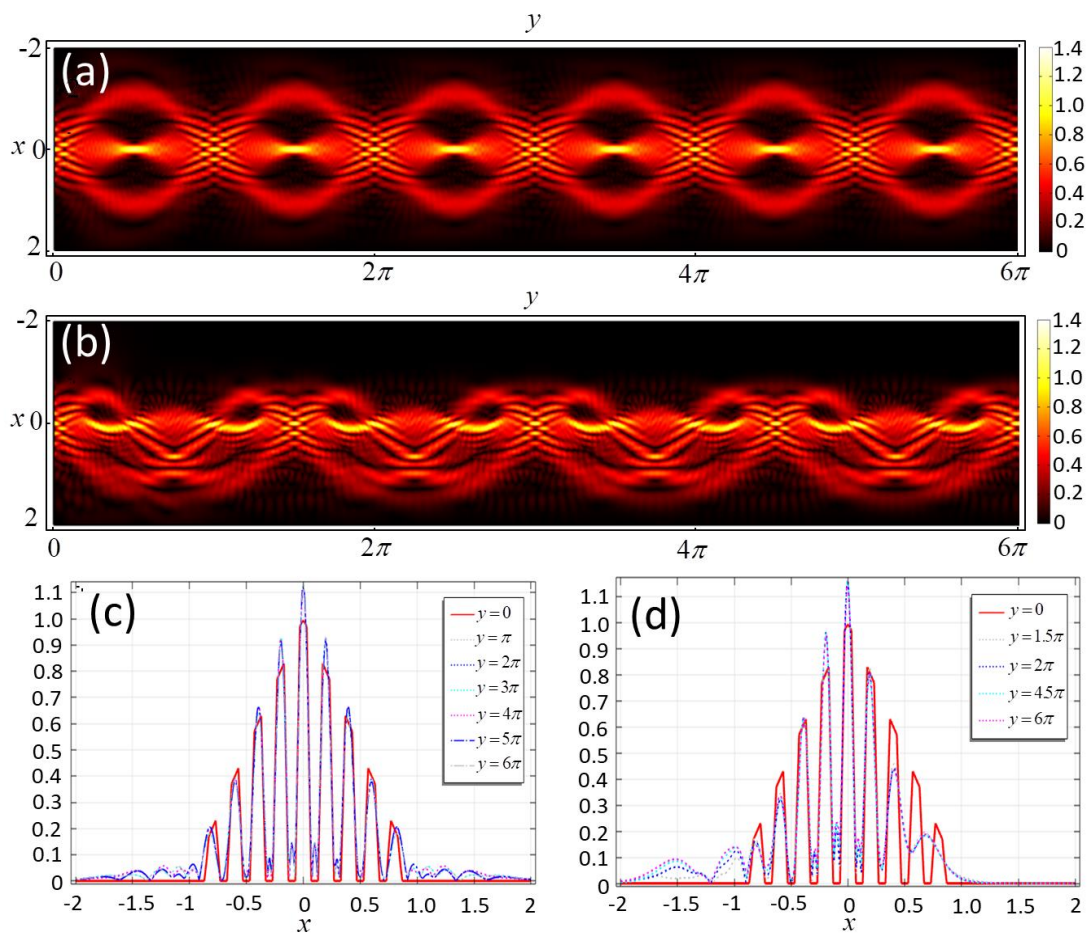

5(d), which demonstrates Talbot effect in this duplex Mikaelian lens. The main difference of the two Talbot effect is that their periods and the symmetry of the field profile are different, which comes from the mirror symmetry of the Mikaelian lens and asymmetry of the duplex Mikaelian lens. In wave optics, duplex Mikaelian lenses and their exponential conformal transformers are not equivalent, which deserves more investigation.

Fig. 5 The Talbot effect in the Mikaelian lens and duplex Mikaelian lens. (a) Electromagnetic field (with unit V/m) out of plane is plotted in the Mikaelian lens of Fig. 1(a). (b) Electromagnetic field is plotted in the duplex Mikaelian lens of Fig. 2(a). The metallic grating source with its magnitude satisfying $abs(1-y)$ is set at $x$ axis; (c) Electromagnetic field out of plane at every half period in the Mikaelian lens; (d) Electromagnetic field out of plane at every period in the duplex Mikaelian lens.

## VI. Conclusion

In summary, we report two new kinds of AOIs in 2D space. One is the duplex Mikaelian lens with any two different positive real number $\{p,q\}$ resulting from continuous translation symmetry described by Lie group $R^1$. The other is duplex Maxwell's fish eye lens with rational number $\{p,q\}$ because of its continuous rotational symmetry described by Lie group $S^1$. A heuristic explanation is given based on exponential conformal mapping. Owing to the asymmetric gradient refractive index, the caustic effect of geometric optics is shown in duplex Mikaelian lenses and duplex Maxwell's fish eye lenses by ray tracing method. Talbot effect of wave optics is also demonstrated in duplex Mikaelian lenses with numeric simulations. The duplex Maxwell's fish eye lens with rational number $\{p,q\}$ should also be valid in 3D space. Our results flourish the family of AOIs and their imaging properties. These new AOIs should also have the counterparts in surface plasmons, acoustics and so on, hence might bring new applications based on asymmetric gradient refractive indexes.

## Acknowledgements

This work was financially supported by the National Natural Science Foundation of China (Grants No. 11874311, No. 11904006, No. 11690033, No. 61425018, No. 11621091, and No. 11704181), National Science Foundation of Anhui Province of China (1908085QA20), National Key R&D Program of China (Grant No. 2017YFA0303702), National Key Research and Development Program of China (Grant No. 2017YFA0205700), the Recruitment Program for Leading Talent Team of Anhui Province (2019-16) and the Fundamental Research Funds for the Central Universities (Grant No. 20720170015). Senlin Liu, Yuze Wu, Yi Yang, and Zichun Zhou are undergraduate students in Xiamen University Malaysia. We thank the support from Prof. Zhong Chen, Prof. Huiqiong Wang, Prof. Yinshui Fang, Mr. Jingfeng Chen, Miss Yuling Zheng, Mr. Jianqing Zhou, Miss Shanshan Lin, Miss Sicen Tao, Miss Ying Chen, Mr. Shan Zhu, Mr. Pengfei Zhao.